\documentclass[aps,superscriptaddress,twocolumn,pre,longbibliography,floatfix]{revtex4-1}

\usepackage{amsmath,amsthm,amsfonts,amssymb,bm,graphicx,color,mathpazo,times, braket}
\usepackage[colorlinks={true}, citecolor={blue}, filecolor={blue}, linkcolor={blue}, urlcolor={blue}]{hyperref}
\usepackage[caption=false]{subfig}
\usepackage{graphicx}
\usepackage{xcolor} 
\usepackage{graphicx}
\usepackage{soul}
\usepackage{float}
\usepackage{tikz}
\usepackage{multirow}
\usetikzlibrary{positioning}
\usepackage{appendix}
\usepackage{amsmath}
\usepackage{enumerate}
\usepackage[caption=false]{subfig}  

\allowdisplaybreaks
\begin{document}
\title{Harnessing Floquet dynamics for selective metrology in few-qubit systems}
\author{Asghar Ullah}
\email{aullah21@ku.edu.tr}
\author{Hasan Mermer}
\email{hmermer20@ku.edu.tr}
\author{Melih Özkurt}
\email{mozkurt20@ku.edu.tr}
\affiliation{Department of Physics, Ko\c{c} University, 34450 Sar\i yer, Istanbul, T\"urkiye}
\author{Igor Lesanovsky}
\affiliation{Institut für Theoretische Physik and Center for Integrated Quantum Science and Technology, Universität Tübingen,
Auf der Morgenstelle 14, 72076 Tübingen, Germany}
\author{\"Ozg\"ur E. M\"ustecapl\i o\u glu}
	\affiliation{Department of Physics, Ko\c{c} University, 34450 Sar\i yer, Istanbul, T\"urkiye}
	\affiliation{T\"UBİTAK Research Institute for Fundamental Sciences (TBAE), 41470 Gebze, T\"urkiye}

\date{\today} 
\begin{abstract}
Periodically driven quantum systems can function as highly selective parameter filters. We demonstrate this capability in a finite-size, three-qubit system described by the transverse-field Floquet Ising model. In this system, we identify a period-doubling (PD) dynamical phase that exhibits a stark asymmetry in metrological sensitivity to the magnetic field applied on the qubits and to the coupling strength between the qubits. The PD phase originates from $\pi$-pairing, where the initial state exhibits strong overlap with $\pi$-paired Floquet eigenstates, leading to robust period-doubled dynamics and enhanced metrological sensitivity. The analysis of quantum Fisher information reveals that the PD regime significantly enhances precision for estimating the Ising interaction strength while simultaneously suppressing sensitivity to the transverse magnetic field. Conversely, non-PD regimes are optimal for sensing the transverse field. This filtering effect is robust for larger system sizes and is quantifiable using experimentally accessible observables, such as magnetization and two-qubit correlations, via the classical Fisher information. Our work shows that distinct dynamical regimes in finite-size Floquet systems can be harnessed for targeted quantum sensing.
\end{abstract}
\maketitle
\section{Introduction}\label{Sec:intro}

Periodically driven quantum systems, described by Floquet theory, offer a platform to realize novel non-equilibrium phases of matter, most notably discrete time crystals (DTCs), which spontaneously break discrete time-translation symmetry under periodic driving~\cite{PhysRevLett.114.251603,RevModPhys.95.031001,Sacha_2018, Else2020, PhysRevLett.109.160401, PhysRevLett.111.250402, PhysRevLett.116.250401, PhysRevB.93.245146, PhysRevX.7.011026, Herper, PhysRevLett.120.180603, PhysRevLett.126.020602, PhysRevA.99.033618, PhysRevLett.130.120403, PhysRevLett.120.110603, PhysRevLett.130.130401, Keßler_2020, Kuroś_2020, PhysRevLett.127.043602}. While such symmetry breaking is forbidden in equilibrium, corroborated by no-go theorems~\cite{PhysRevLett.111.070402, PhysRevLett.114.251603}, it becomes possible in driven systems when energy absorption is suppressed, e.g., by many-body localization~\cite{PhysRevLett.117.090402} or prethermalization~\cite{PhysRevB.95.014112}. It is crucial to emphasize that true time-crystalline order is a many-body phenomenon, requiring a thermodynamic limit to exhibit robust, spontaneous symmetry breaking~\cite{PhysRevLett.118.030401,RevModPhys.95.031001}. Nevertheless, the hallmark signatures of DTCs, specifically a subharmonic response such as period-doubling (PD), can manifest even in minimal, finite-size quantum systems~\cite{deshmukh2025observation}. Experimental realizations of such driven phases have been reported~\cite{PhysRevLett.127.043602,Phattam,Wu2024, Kongkhambut_2024, PhysRevA.109.063317}, and their dynamics have also been investigated for quantum metrological applications~\cite{Montenegro2023,PhysRevLett.132.050801,PhysRevA.109.L050203, PhysRevB.111.125159, PhysRevB.111.024315}. Such small systems, accessible on near-term quantum devices, can serve as ideal platforms not for studying true phase transitions, but for harnessing these distinctive dynamical regimes for functional purposes.

Quantum metrology aims to estimate physical parameters with high precision~\cite{Giovannetti2011,Sidhu, MONTENEGRO20251}. In systems influenced by several parameters, the estimation of one parameter can be influenced by fluctuations in others, making careful selection of the sensing regime essential. Periodically driven quantum systems can provide an approach to address this challenge, as their dynamical phases can be tuned to enhance sensitivity to a target parameter while suppressing responses to others. The central thesis of this work is that these dynamical regimes can be exploited to create a highly selective quantum sensor. Our analysis in the present work does not rely on the thermodynamic limit but focuses on relevant finite-size systems. In this regime, Floquet dynamics act as an effective filter for parameter sensitivity, enabling selective enhancement or suppression of metrological precision depending on the operating point. We demonstrate this principle using a minimal three-qubit Ising model subject to a transverse field, where the dynamical phase serves as a functional switch for metrology. By driving the system into a PD regime, it becomes a powerful parameter filter: its sensitivity to the Ising interaction strength is dramatically enhanced. At the same time, its response to the transverse magnetic field is simultaneously suppressed. This ability to actively select which parameter to sense is a key advantage for quantum metrology in noisy environments.

We establish this metrological trade-off by first calculating the quantum Fisher information (QFI), which provides the ultimate, measurement-independent bound on precision. More importantly, to connect our findings to practical applications, we demonstrate that this selective sensitivity is fully captured by the classical Fisher information (CFI) for experimentally accessible observables, namely the total magnetization and two-qubit correlations. These results confirm that the metrological advantage is not a theoretical abstraction but can be extracted with standard measurement techniques on platforms like trapped ions~\cite{Zhang2017} or superconducting qubits~\cite{Mi2022,Frey2022}. Furthermore, we show that these features persist for larger system sizes, demonstrating the robustness of the observed behavior. Our work positions small-scale Floquet systems as powerful, tunable tools for near-term quantum sensing.

The rest of the paper is organized as follows. In Sec.~\ref{model}, we present our minimal three-qubit Floquet model. We discuss the dynamical regimes in Sec.~\ref{dynamics}. We explore the metrological applications of our system for the estimation of magnetic field and coupling strength using QFI and CFI in Sec.~\ref{sec:metrology}. We summarize our work in Sec.~\ref{conclusion}. We present the QFI curvature and scaling analysis in PD and non-PD regimes in Appendix~\ref{app:curvature}, discuss the different dynamical regimes in Appendix~\ref{dynamic}, and show the results for larger system sizes in Appendix~\ref{largerSpins}.

\section{The Model}\label{model}
\begin{figure}
    \centering
    \includegraphics[width=0.95\linewidth]{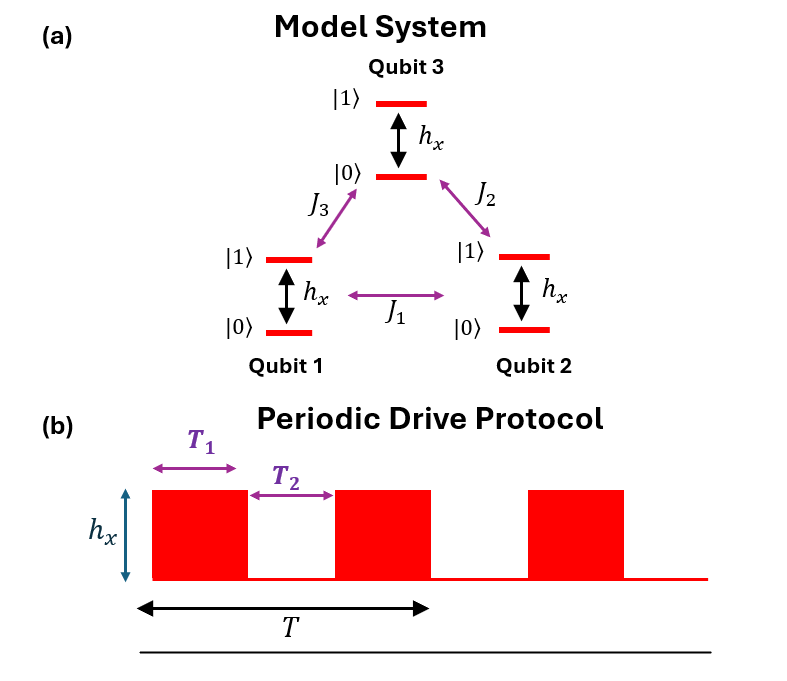}
    \caption{(a) Sketch of the three-qubit model system. Qubits interact via nearest-neighbor Ising coupling strengths ($J_1$, $J_2$, $J_3$) and are driven by a transverse magnetic field $h_x$. (b) The dynamics are generated by a two-step Floquet protocol: each period $T=T_1+T_2$ consists of a pulse of amplitude $h_x$ during $T_1$, which is followed by Ising evolution for duration $T_2$.}
    \label{fig:model}
\end{figure}
We investigate the non-equilibrium dynamics of a minimal, finite-qubit system under periodic driving, modeled by the transverse-field Ising Hamiltonian as depicted in Fig.~\ref{fig:model}. Our choice of a three-qubit system is deliberate: it serves as a minimal model to explore the emergence of complex dynamical signatures, such as PD, in a setting that is directly relevant to near-term quantum hardware. While we focus on this few-body regime, we also verify in Appendix~\ref{largerSpins} that our central metrological findings are robust and persist in larger systems, demonstrating their scalability. The system is evolved via a Floquet protocol, where each driving period \(T\) is composed of two sequential stages:
\begin{enumerate}
    \item A global transverse magnetic field pulse of strength \(h_x\) applied along the \(x\)-axis for a duration \(T_1\).
    \item An Ising interaction of strength \(J\) applied along the \(z\)-axis for a duration \(T_2 = T - T_1\).
\end{enumerate}

The total Floquet unitary describing the evolution over one period is constructed as:
\begin{equation}
    \hat{\mathrm{U}}_F = e^{-i \hat{H}_z T_2} e^{-i \hat{H}_x T_1},
\end{equation}
where the transverse-field Hamiltonian with strength $h_x$ is:
\begin{equation}
    \hat{H}_x = h_x \left( \hat{\sigma}^1_x + \hat{\sigma}^2_x + \hat{\sigma}^3_x \right),
\end{equation}
and the Ising interaction Hamiltonian under periodic boundary conditions is:
\begin{equation}
    \hat{H}_z = J_1\hat{\sigma}^1_z \hat{\sigma}^2_z + J_2\hat{\sigma}^2_z \hat{\sigma}^3_z+J_3\hat{\sigma}^1_z \hat{\sigma}^3_z, 
\end{equation}
where $J_{\alpha}$($\alpha=1,2,3$) characterize the coupling strength between the spins and $\hat{\sigma}^i_\alpha$ ($\alpha=x,y,z$) are the Pauli spin-1/2 operators acting on the i-th spin. Although the Hamiltonian is written in the conventional spin language, in the remainder of this work, we refer to the degrees of freedom as qubits, in line with their realization on current quantum hardware~\cite{Simon, Wolf, Johnson2011, Zhang2017}. Throughout this work we fix the driving protocol as $T_1 = 0.5T$ and $T_2 = T - T_1$, 
and assume uniform couplings $J_1=J_2=J_3=J$.

\section{Dynamical regimes}\label{dynamics}
\subsection{Period doubling}

\begin{figure}[t!]
    \centering
    \subfloat[]{
    \includegraphics[scale=0.34]{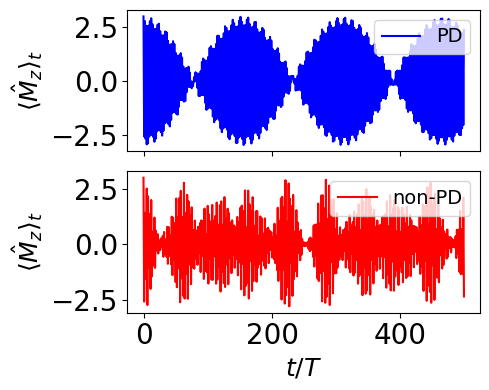}}
    \subfloat[]{
    \includegraphics[scale=0.35]{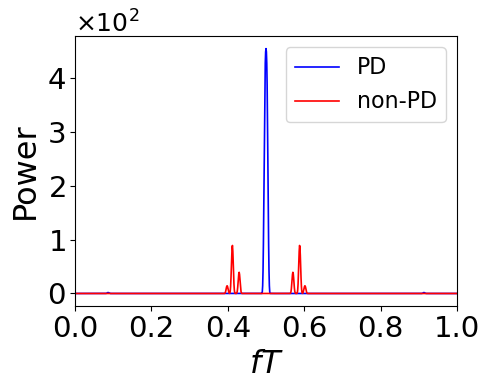}}
    \caption{(a) Average magnetization $\langle \hat{M}_z \rangle_t$ at stroboscopic times \( t = nT \), where \( n \in \mathbb{Z} \),  for PD (blue) and non-PD (red) regimes. The PD regime shows robust PD oscillations, while the non-PD case shows trivial dynamics. (b) Power spectrum of the magnetization dynamics, which shows a strong subharmonic response in the PD regime. For the PD regime, we use $h_xT=2.6$ and $JT=1.57$, while for the non-PD regime, we use $h_xT=2.6$ and $JT=0.1$.
}
    \label{fig:Mz+Power+QE}
\end{figure}

To identify the PD signature, we simulate the dynamics starting from the initial state \(\ket{000}\), polarized along the \(z\)-axis, and compute the average magnetization along \(z\) at stroboscopic times $t = nT$ as
\begin{equation}
    \langle \hat{M}_z \rangle_{t=nT} =
    \bra{000}
    \big( \hat{\mathrm{U}}_F^\dagger \big)^{n}
    \left(\sum_{i=1}^{3} \hat{\sigma}_z^i \right)
    \big( \hat{\mathrm{U}}_F \big)^{n}
    \ket{000},
\end{equation}
where \( \hat{\mathrm{U}}_F \) denotes the Floquet operator for one driving period. We focus on the $z$-magnetization, which is aligned with the Ising interaction and the Floquet pulse and therefore directly captures the PD response. This quantity serves as an order parameter reflecting the collective qubit dynamics of the system. In the PD phase, \(\langle \hat{M}_z \rangle_t\) exhibits oscillations with a period of \(2T\), whereas in the trivial phase, such as non-PD, it oscillates with a period \(T\) or quickly decays to a steady value, shown in Fig.~\ref{fig:Mz+Power+QE}(a).

To characterize this behavior quantitatively, we perform a discrete Fourier transform of the magnetization time series after discarding transient effects (first 50 periods) and search for a prominent spectral peak at frequency \(f =1 /2T\). The presence of such a subharmonic peak signals the PD phase. Figures~\ref{fig:Mz+Power+QE}(a) and~\ref{fig:Mz+Power+QE}(b) illustrate the magnetization dynamics and their corresponding power spectra for the selected PD ($h_xT=2.6$, $JT=1.57$) and non-PD ($h_xT=2.6$, $JT=0.1$) regimes. In the PD regime, we can see that the magnetization \( \langle \hat{M}_z \rangle_t \) exhibits clear PD oscillations, consistent with a time-crystalline-like response. This PD is further supported by a prominent subharmonic peak at approximately $f = 1/2T$ in the power spectrum shown in Fig.~\ref{fig:Mz+Power+QE}(b). Conversely, in the non-PD regime, \( \langle \hat{M}_z \rangle_t \) displays irregular oscillations with weak or transient subharmonic features (red curve in Fig.~\ref{fig:Mz+Power+QE}(a)), and its power spectrum reveals subharmonic peaks at frequencies different from $1/2T$, unlike the PD regime where a dominant subharmonic at $1/2T$ is clearly observed (Fig.~\ref{fig:Mz+Power+QE}(b)). In addition, the Fourier spectrum of the real-valued magnetization signal is symmetric around the Nyquist frequency of $f=1/2T$.
\subsection{PD phase diagram and $\pi$-pairing}
\begin{figure}[t!]
    \centering
    \subfloat[]{
    \includegraphics[scale=0.28]{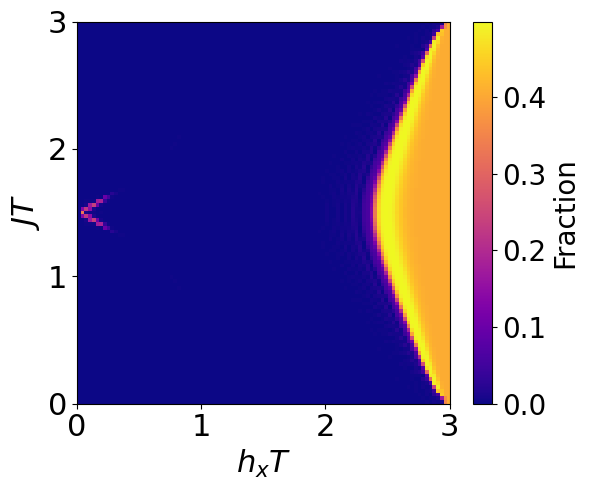}}
    \subfloat[]{
        \includegraphics[scale=0.28]{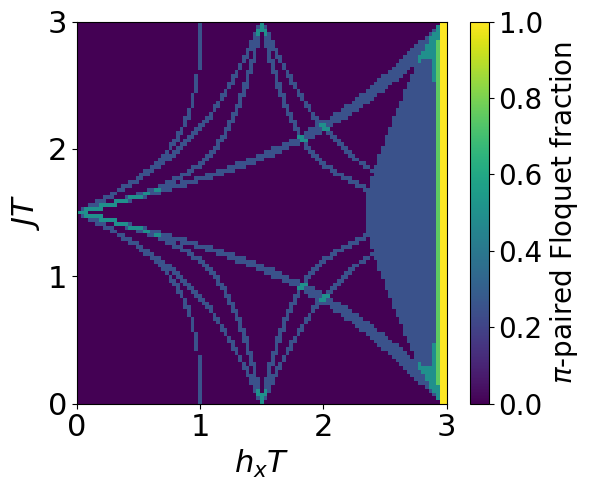}}
    \caption{(a) Phase diagram of the relative subharmonic spectral weight, quantifying the fraction of oscillatory power contained in the subharmonic mode $f=1/2T$. Regions of high spectral weight indicate strong PD dynamics. (b) Corresponding fraction of $\pi$-paired Floquet eigenstates, obtained from the quasienergy spectrum of the Floquet operator $\mathrm{\hat{U}_F}$. The presence of pairs of Floquet states with quasienergy separation close to $\pi/T$ underlies the emergence of PD behavior in finite systems.}
    \label{fig:PD_pi_pair}
\end{figure}

To quantitatively map the phase diagram of the PD regime, we analyze the stroboscopic magnetization dynamics, $\hat{M}_z(nT)$. A defining signature of the PD phase is the emergence of a subharmonic response at frequency $f = 1/2T$, which can be identified by the Fourier transform of the time series. To construct a sensitive diagnostic, it is essential to separate the oscillatory component of the dynamics from its static polarization. We therefore isolate the purely dynamic part of the signal as 
\begin{equation}
    \delta \hat{M}_z(nT) = \hat{M}_z(nT) - \langle \hat{M}_z \rangle,
\end{equation}
where $\langle \hat{M}_z \rangle$ denotes the time-averaged magnetization over the stroboscopic evolution, after discarding an initial transient period. From the power spectrum of this dynamic signal, $\delta M_z(nT)$, we define the \textit{relative subharmonic spectral weight} as our diagnostic (see Fig.~\ref{fig:PD_pi_pair}(a)). This quantity is given by the ratio of the spectral power at the subharmonic frequency $f = 1/2T$ to the total power summed over all non-zero frequencies. The resulting measure, bounded between 0 and 1, quantifies the fraction of the system’s total oscillatory power contained in the subharmonic mode. A value approaching unity indicates that the dynamics are almost entirely governed by a single PD frequency, while smaller values signify that the subharmonic response coexists with other dynamic fluctuations. This metric thus provides a quantitative criterion for distinguishing between regimes of coherent PD and more complex, multifrequency dynamics in our finite-sized system.
To highlight the underlying mechanism, we analyze the quasienergy spectrum of the Floquet operator, as shown in Fig.~\ref{fig:PD_pi_pair}(b). By diagonalizing the unitary $\hat{\mathrm{U}}_F$, we identify pairs of eigenstates whose quasienergies differ by approximately \( \pi/T \), defining the fraction of $\pi$-paired states and their overlap with the initial state. In DTCs, the eigenstates of the Floquet operator appear in pairs whose quasienergies (i.e., phases modulo $2\pi/T$) differ by $\pi/T$~\cite{PhysRevB.111.184308}. A superposition of such paired states reproduces itself after two driving periods \(2T\), thereby giving rise to PD. Because \(\pi\)-pairing is a spectral property of the Floquet operator, it can also occur in small systems such as our three-qubit ring, serving as the essential mechanism underlying the PD observed in Fig.~\ref{fig:PD_pi_pair}.

However, there are several important distinctions from the conventional DTC scenario: We consider an initial state that only partially overlaps with the Floquet eigenstates. A trivial \(\pi\)-pairing can be enforced by fixing both the spin-flip (\(\hat{\sigma}_x\)) and pairwise Ising interaction terms to \(\pi/2\); however, in our study aimed at sensing applications, we do not fix \(h_x\) and \(J\) to exact \(\pi\)-pairing values. Finally, we also examine \emph{imperfect} PD points in addition to ideal PD.  

As a result, only a fraction of the eigenstates are \(\pi\)-paired with higher spectral weight, and this fraction varies across the \((h_xT, JT)\) parameter space. In summary, while \(\pi\)-pairing remains the underlying mechanism of PD or approximate PD dynamics in our finite system, the dependence on the initial state, system size, and parameters implies that PD here lacks the robustness to perturbations or scaling that characterizes a genuine DTC. While DTCs are defined as robust phases of matter protected against disorder and imperfections, the PD behavior discussed here does not necessarily satisfy all such stability criteria. In particular, we do not claim robustness against disorder or perturbations in this finite-size system. Nevertheless, such finite-size and partially paired regimes still exhibit transient or quasi-periodic PD signatures, providing a minimal model to explore the onset of DTC-like dynamics and their connection to metrological precision. 
We emphasize that the results presented here pertain to closed, finite-size Floquet unitary dynamics. In this setting, observables can exhibit long-lived oscillations, beats, and revivals arising from discrete quasienergy spacings; therefore, the thermodynamic notion of irreversible Floquet heating toward an infinite-temperature state is not directly addressed. Consequently, the PD dynamics and $\pi$-pairing identified in this work are employed to characterize distinct finite-size dynamical regimes, rather than to establish prethermal lifetimes or quantify heating suppression. A systematic analysis of energy absorption, heating rates, and possible prethermal behavior becomes essential when extending these ideas to larger many-body systems or to experimentally relevant open-system settings. Investigating such effects, including their dependence on the parameters $(h_x T, J T)$, constitutes an important direction for future work.

\section{Metrological applications}\label{sec:metrology}
Having established the existence of distinct dynamical regimes, we now investigate their metrological utility. In this section, we test our central hypothesis: that these phases can be harnessed for selective parameter estimation. We evaluate the ultimate precision limits for sensing the magnetic field $h_x$ and interaction strength $J$ using the QFI, and then connect these limits to practical measurements via the CFI by measuring all quantities in units of $T$.
\subsection{Quantum parameter estimation theory}
Quantum parameter estimation theory aims to infer an unknown quantity $\theta$ encoded in a quantum state $\rho(\theta)$. For a given measurement described by a set of positive operator valued measures (POVM) $\{\Pi_s\}$, the Cramér-Rao bound sets a lower bound on the estimation of the parameter $\theta$, such as~\cite{PhysRevLett.72.3439}
\begin{equation}
    \mathrm{Var}(\theta)\ge\frac{1}{mF_C(\theta)},
\end{equation}
where $\mathrm{Var}(\theta)$ is the variance, $m$ is the number of repeated measurements, and $F_C(\theta)$ denotes the CFI, which for a probability distribution $p(s|\theta)=\text{Tr}(\Pi_s\rho(\theta))$ is given by~\cite{Fisher1992, Fisher_1925}
\begin{equation}
    F_C(\theta)=\sum_s p(s|\theta)\bigg[\frac{\partial_\theta p(s|\theta)}{p(s|\theta)}\bigg]^2.
\end{equation}
Optimizing over all possible POVMs defines the QFI, given by
\begin{equation}
F_Q(\theta) = \max_{\{\Pi_s\}} F_C(\theta),    
\end{equation}
which sets the ultimate bound on the achievable precision. Hence, the quantum Cramér-Rao bound gives~\cite{paris2004quantum, PhysRevLett.72.3439, helstrom1969quantum}:
\begin{equation}
   \mathrm{Var}(\theta)\ge\frac{1}{mF_C(\theta)}\ge\frac{1}{mF_Q(\theta)},
\end{equation}
For pure quantum states, \(\rho(\theta) = |\psi(\theta)\rangle\langle \psi(\theta)|\), the QFI can be evaluated using the following simplified form ~\cite{Fisher1992}
\begin{equation}\label{QFI:pure}
    F_Q(\theta)=4\big[ \langle \partial_\theta \psi(\theta) | \partial_\theta \psi(\theta) \rangle - |\langle \psi(\theta) | \partial_\theta \psi(\theta) \rangle|^2 \big].
\end{equation}
The QFI is used to analyze the control parameter $\theta$, which in our case is the coupling strength $J$ and transverse magnetic field $h_x$.
\begin{figure}[t!]
    \centering
    \subfloat[ Estimation of $h_x$]{
    \includegraphics[scale=0.265]{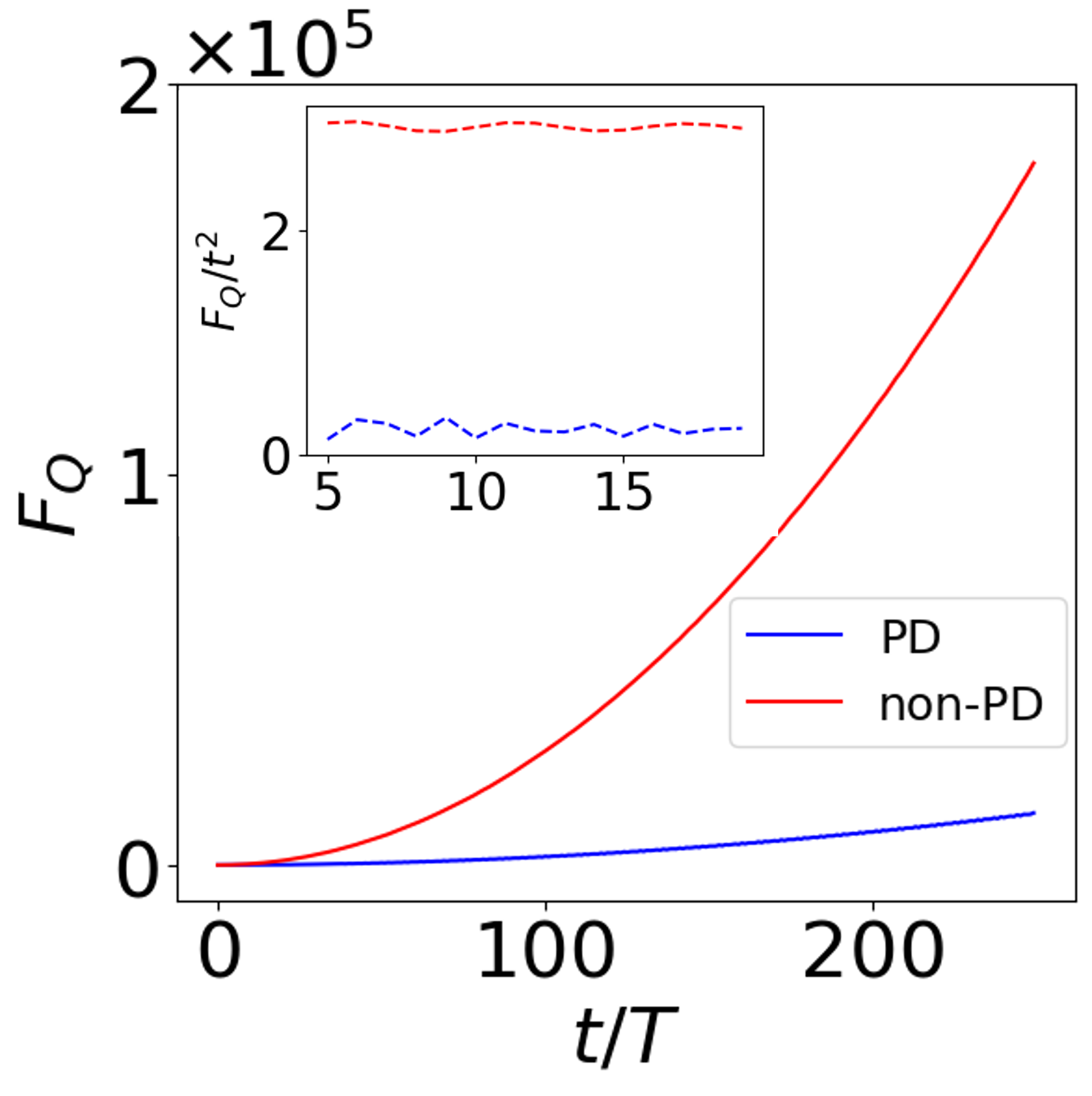}}
    \subfloat[Estimation of $J$]{
        \includegraphics[scale=0.265]{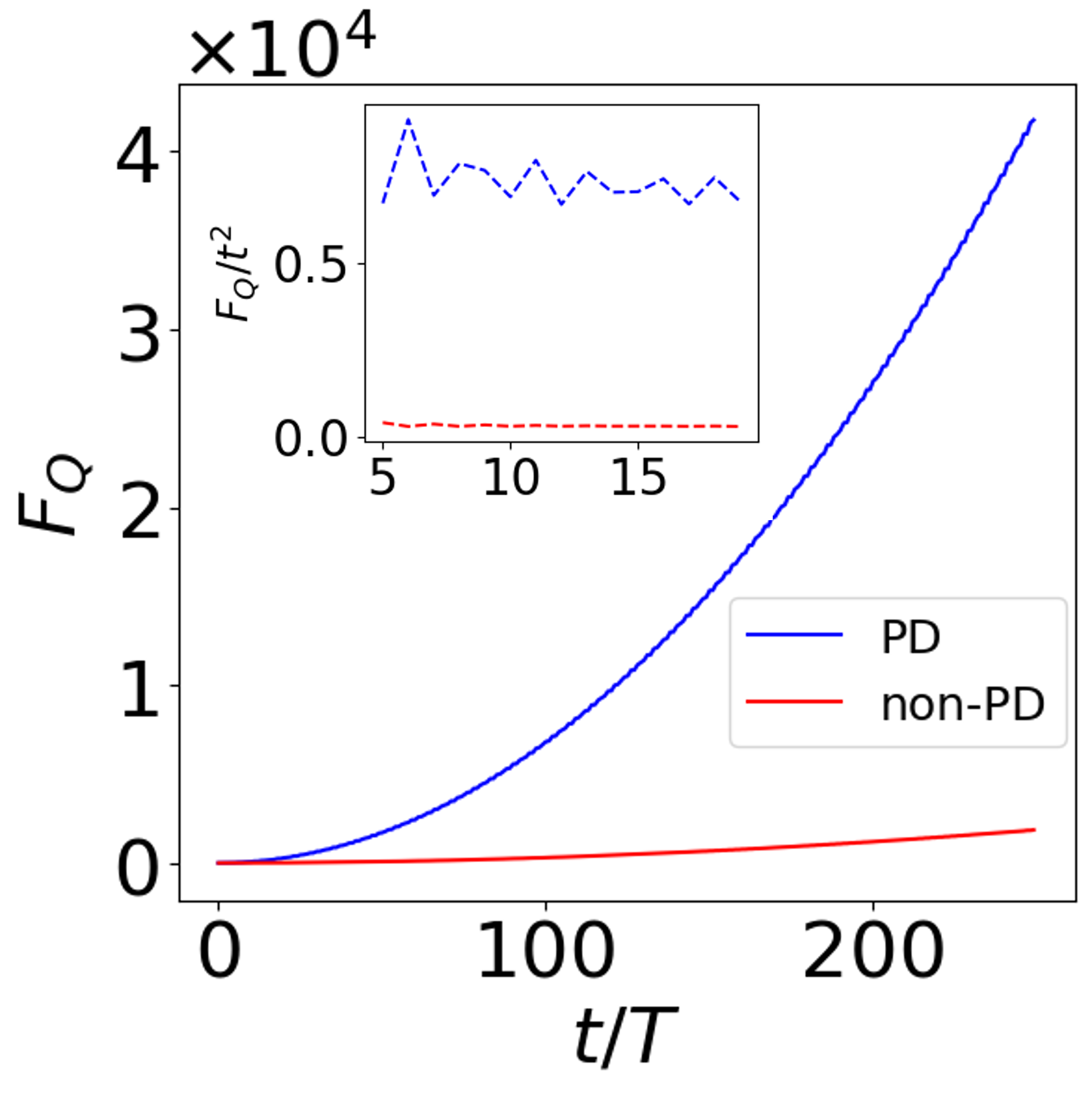}}
    \caption{(a) QFI dynamics for the estimation of the transverse field $h_x$, calculated from the Floquet-evolved states, as a function of stroboscopic time $t/T$. The QFI is plotted for the selected PD (blue) and non-PD (red) parameter sets. For the PD regime, we use $h_xT=2.6$ and $JT=1.57$, while for the non-PD regime, we use $h_xT=2.6$ and $JT=0.1$. (b) QFI dynamics for estimating the interaction strength $J$,  calculated from the Floquet-evolved states, as a function of stroboscopic time $t/T$.  For the PD regime, we use $h_xT=2.6$ and $JT=1.57$, while for the non-PD regime, we use $h_xT=0.1$ and $JT=1.57$. Insets in both panels show $F_Q/t^2$ at short times to highlight the absence of quadratic scaling at early times.}
    \label{fig:QFI}
\end{figure}
\subsection{Estimation of transverse magnetic field $h_x$ and interaction strength $J$}
Throughout this work, the QFI is employed as a theoretical benchmark to quantify the ultimate precision limits for estimating the parameters $h_x$ and $J$, assuming optimal quantum measurements. Rather than fixing a particular measurement operator, we compute the QFI, which represents the maximum achievable precision attainable over all possible POVMs. The extent to which this bound can be saturated in practice depends on the specific measurement protocol, a point we address in section~\ref{sec:CFI}.

For consistency with the earlier discussion of dynamical regimes, we initialize the system in the fully $z$-polarized state $\ket{000}$. For the estimation of the magnetic field $h_x$, we focus on parameters from the phase diagram in Fig.~\ref{fig:Mz+Power+QE}(a), where the subharmonic spectral weight is large, corresponding to robust PD dynamics, and are given as

\begin{itemize}
    \item PD regime: \(h_xT = 2.6\), \(JT = 1.57\),
    \item non-PD regime: \(h_xT = 2.6\), \(JT = 0.10\).
\end{itemize}

These points are chosen such that the transverse field strength \( h_x \) remains fixed across both regimes, allowing a direct comparison of sensing performance while varying only the interaction strength, which governs the transition between the PD and non-PD phases.

The QFI for estimation of $h_x$ is plotted as a function of time (in units of driving periods $T$) for the selected PD (blue) and non-PD parameter sets in Fig.~\ref{fig:QFI}(a). The non-PD regime exhibits faster QFI growth compared to the PD regime, indicating a higher sensitivity for $h_x$ estimation in the trivial non-PD phase. The QFI growth follows the quadratic scaling $F_Q\propto t^2$, as detailed in Appendix~\ref{app:curvature}. To assess deviations from this scaling at short evolution times, we additionally show the normalized quantity $F_Q/t^2$ in the inset of Fig.~\ref{fig:QFI}(a).

This difference can be understood from the roles the parameters play in sustaining the dynamics. The PD phase is stabilized by the Ising interaction term $\hat{H}_z$, which preserves the system's spin-flip ($\mathbb{Z}_2$) symmetry. The transverse field $\hat{H}_x$, however, explicitly breaks this symmetry. Consequently, the coherent PD dynamics are inherently robust against the symmetry-breaking perturbation ($h_x$), making the system a poor sensor for it. This robustness is a hallmark of symmetry-protected dynamics~\cite{PhysRevB.111.184308, PhysRevLett.116.250401, PhysRevLett.127.043602}. In essence, the PD regime functions as a parameter filter: it is highly responsive to changes in the parameter that stabilizes it ($J$), while suppressing sensitivity to the parameter that perturbs it ($h_x$). This selective insensitivity is precisely why the QFI for $h_x$ is suppressed in the PD phase.

As a result, PD regimes act as a parameter filter, being highly responsive to symmetry-preserving changes such as $J$, while suppressing the metrological sensitivity to symmetry-breaking perturbations like $h_x$. This selectivity, captured in the QFI behavior, suggests that PD regimes are better suited for interaction-based sensing, where enhanced precision is linked to the role of $J$ in sustaining subharmonic stability. To complement this analysis, we next investigate the estimation of the interaction strength, where the PD regime is expected to exhibit enhanced sensitivity due to its structural dependence on $J$.

For estimation of the coupling strength $J$, we select the following parameters for comparison:

\begin{itemize}
    \item PD regime: \(h_xT = 2.6\), \(JT = 1.57\),
    \item non-PD regime: \(h_xT = 0.1\), \(JT = 1.57\).
\end{itemize}

These points are chosen such that $J$ remains fixed across both regimes, while we vary the value of $h_x$, which governs the distinction between the PD and non-PD regimes.

We present our results for the estimation of the interaction strength \( J \) in both the PD and non-PD regimes. As shown in Fig.~\ref{fig:QFI}(b), the QFI grows rapidly in the PD regime, following a superlinear trend, while it remains significantly suppressed in the non-PD regime. Likewise, the inset in Fig.~\ref{fig:QFI}(b) shows $F_Q/t^2$ to reveal the short-time behavior earlier than the quadratic regime. This pronounced contrast highlights the enhanced sensitivity of the PD phase to variations in \( J \), consistent with its structural role in stabilizing subharmonic dynamics. In contrast, the non-PD regime lacks such symmetry-protected coherence, resulting in much weaker metrological performance.
\begin{figure}[t!]
    \centering
    \subfloat[Estimation of $h_x$]{
    \includegraphics[scale=0.42]{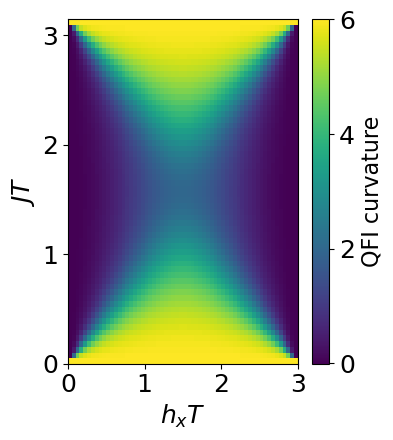}}
    \subfloat[Estimation of $J$]{
    \includegraphics[scale=0.42]{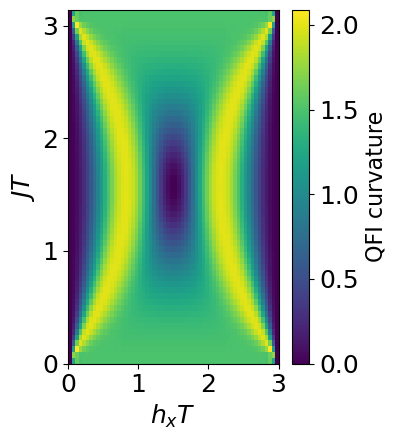}}
    \caption{Phase diagrams of QFI curvature $d^2F_Q/dt^2$ (Eq.~\eqref{QFI_curve}) as a function of interaction strength $J$ and transverse field $h_x$ for estimation of $h_x$ (a) and $J$ (b). The rest of the parameters are the same as in Fig.~\ref{fig:Mz+Power+QE}.}
    \label{fig:Pd_vs_nPD}
\end{figure}

To quantify the scaling of metrological sensitivity, we consider the curvature of the QFI with respect to time, defined as
\begin{equation}\label{QFI_curve}
\kappa_\alpha(h_x,J) \equiv \frac{d^2 F_Q(\alpha)}{dt^2}, \qquad \alpha \in \{h_x, J\}.
\end{equation}
We consider $\kappa_\alpha$ because $F_Q(t)$ grows approximately quadratically at long times, so the curvature provides a simple yet meaningful measure of the rate at which sensitivity to the parameter $\alpha$ changes. This allows us to compare metrological performance across different regions of the $(h_xT,JT)$ space. 

Figure~\ref{fig:Pd_vs_nPD} presents the phase diagrams of the QFI curvature 
$\kappa_\alpha$ for the estimation of $h_x$ (a) and $J$ (b), obtained as 
a function of both $h_x$ and $J$. This quantity isolates the leading quadratic 
growth of the QFI with time as given in Fig.~\ref{fig:QFI}, thereby providing a measure of the intrinsic parameter sensitivity of the system. For the estimation of $h_x$ (Fig.~\ref{fig:Pd_vs_nPD}(a)), the curvature is maximized outside the PD phase, in agreement with Fig.~\ref{fig:QFI}(a) where the non-PD regime showed enhanced QFI growth for estimation of $h_x$. In contrast, for the estimation of $J$ (Fig.~\ref{fig:Pd_vs_nPD}(b)), the curvature is strongly enhanced near the PD regime, which is consistent with Fig.~\ref{fig:QFI}(b). We observe that the Floquet phase diagram is not symmetric across the full $h_xT$--$JT$ domain, but rather exhibits an arc on the right side in the strong $h_x$ regime, as shown in Fig.~\ref{fig:PD_pi_pair}(a). In practice, the observed structure in Fig.~\ref{fig:PD_pi_pair}(a) depends on the overlap of the chosen initial state with the Floquet eigenstates. Therefore, when estimating the interaction strength, it is important to select values of $h_x$ corresponding to the PD points (see Fig.~\ref{fig:Pd_vs_nPD}(b)), which lie in the region $h_xT > 2$. While the PD points in this region do not yield the absolute maximum precision, the QFI remains relatively high, allowing one to control the estimation precision by avoiding other $h_x$ values.

These results highlight that PD dynamics amplify sensitivity to the interaction strength $J$, while the non-PD regime favors estimation of the transverse field $h_x$. The PD phase is robust yet consistently poor for $h_x$ estimation, offering moderate precision for $J$ without dropping to very low performance, whereas the non-PD phase can yield higher sensitivity but with less predictable outcomes. Therefore, when estimating $h_x$, one should carefully choose $J$ to avoid parameter regions exhibiting PD behavior, since the maximum QFI lies in the non-PD regime. This can be understood from the Floquet spectrum: in the PD regime, a quasienergy splitting near a subharmonic bifurcation makes the dynamics highly nonlinear in $J$, enhancing its QFI, while the approximate time-crystalline rigidity suppresses the response to a uniform field $h_x$, explaining why $J$ emerges as the more sensitive parameter. Because our analysis focuses on closed, finite-sized Floquet unitary dynamics, the metrological performance discussed here is governed by the structure of the parameter space in $h_xT$ and $hT$, rather than by finite prethermal lifetimes associated with many-body heating.

\subsection{Classical Fisher information and choice of observables}\label{sec:CFI}
In general, the estimation task depends on an unknown parameter, and the corresponding optimal measurement basis may be highly entangled and experimentally inaccessible. It is therefore desirable to assess the estimation performance with sub-optimal yet feasible measurements. While the QFI sets the ultimate precision limit via the Cramér–Rao bound, attaining this bound requires projective measurements in the eigenbasis of the symmetric logarithmic derivative, which are often impractical. In practice, one therefore considers the CFI, which quantifies the sensitivity of specific, experimentally available measurement outcomes. For a measurement associated with an observable $\hat{X}$, the corresponding CFI can be written as~\cite{PhysRevLett.125.080402}
\begin{equation}\label{CFI}
F_C[\hat{X}]=\frac{1}{\langle\Delta\hat{X}^2\rangle}\Big(\frac{\partial\langle\hat{X}\rangle}{\partial \theta}\Big)^2,
\end{equation}
where $\theta$ is the parameter to be estimated (either $h_x$ or $J$), $\langle \hat{X} \rangle$ is the expectation value of $\hat{X}$, and $\langle\Delta \hat{X}^2\rangle$ is its variance.
\begin{figure}[t!]
   \centering
   \subfloat[Estimation of $h_x$]{
        \includegraphics[scale=0.42]{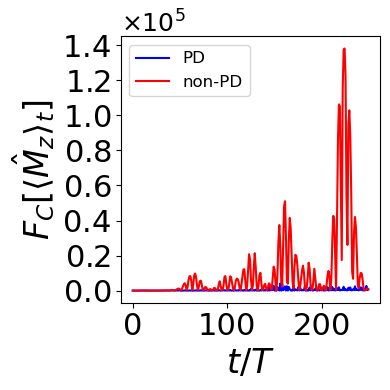}}
        \subfloat[Estimation of $J$]{
        \includegraphics[scale=0.42]{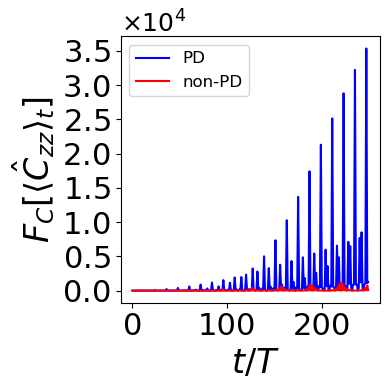}}
    \caption{\textbf{(a)} CFI as a function of time $t/T$ for estimation of magnetic field $h_x$ using total magnetization $\hat{M}_z$ as an observable for PD (blue) and non-PD (red) regimes. For the PD regime, we use $h_xT=2.6$ and $JT=1.57$, while for the non-PD regime, we use $h_xT=2.6$ and $JT=0.1$, \textbf{(b)} CFI as a function of time $t/T$ for estimation of interaction strength $J$ using the two-point correlation function $\hat{C}_{zz}$ as an observable. For the PD regime, we use $h_xT=2.6$ and $JT=1.57$, while for the non-PD regime, we use $h_xT=0.1$ and $JT=1.57$.}
    \label{fig:CFI}
\end{figure}

In our periodically driven three-qubit system, we consider different observables for the calculation of CFI. For the estimation of $h_x$, we can use the total magnetization along the $z$ direction, such that

\begin{equation}
\hat{M}_z = \sum_{i=1}^{3} \hat{\sigma}_z^{i},
\end{equation}
which are directly sensitive to changes in the transverse field $h_x$. For the estimation of the coupling strength $J$, we employ a two-body correlation operator along the $z$-direction, which serves as the standard and experimentally accessible measurement that corresponds to the classical measurement. Specifically, for a chain of three qubits, the operator takes the form
\begin{equation}
\hat{C}_{zz} = \sum_{i<j} \hat{\sigma}_z^{i} \hat{\sigma}_z^{j}
= \hat{\sigma}_z^{1}\hat{\sigma}_z^{2} + \hat{\sigma}_z^{2}\hat{\sigma}_z^{3},
\end{equation}
which captures the nearest-neighbor qubit–qubit correlations induced by the coupling strength $J$. Importantly, this observable is experimentally accessible in state-of-the-art platforms such as superconducting qubits~\cite{Mi2022} and trapped ions~\cite{PhysRevLett.109.163001, Zhang2017}. Figure~\ref{fig:CFI} shows the CFI as a function of time for the estimation of magnetic field strength $h_x$ and interaction strength $J$. For the estimation of $h_x$, we use the magnetization operator as the measurement observable, setting $\hat{X} = \hat{M}_z$ in Eq.~\eqref{CFI} and calculating the CFI in both the PD and non-PD regimes. For the estimation of $h_x$, the CFI is nearly zero at the start of evolution time $T$ and gradually increases. It shows the highest value in the non-PD regime, while the CFI for the PD regime is very small, shown in Fig.~\ref{fig:CFI}(a). The oscillations in the CFI arise from the finite size of the system. In small systems, magnetization naturally oscillates, and these finite-size oscillations are reflected in the CFI dynamics. In contrast, for the estimation of $J$, we take the two-point correlation function as the measurement observable, setting $\hat{X} = \hat{C}_{zz}$ in Eq.~\eqref{CFI}. In this case, the CFI is higher in the PD regime with much stronger oscillations when the two-point correlation function is used as an observable (see Fig.~\ref{fig:CFI}(b)). Meanwhile, it remains negligible in the non-PD regime.


\section{Conclusion}\label{conclusion}

We have explored the interplay between non-equilibrium dynamics and quantum metrology in a minimal, three-qubit system. We found that by periodically driving the system, one can induce distinct dynamical regimes with markedly different metrological properties. We have shown that in certain parameter windows, a phase exhibiting PD oscillations---a signature reminiscent of time-crystalline dynamics---can function as a selective parameter filter. Specifically, our results based on the QFI indicate that this PD regime enhances sensitivity to the Ising interaction strength while simultaneously suppressing sensitivity to the transverse magnetic field, a behavior that is inverted in the non-PD regime.

By identifying the boundaries between PD and non-PD regimes in the Floquet phase diagram, one can strategically choose the initial conditions and system parameters to operate within the desired dynamical phase. In particular, testing whether the system lies in the PD regime provides a clear route to achieving high-contrast sensitivity to the interaction strength 
$J$, accompanied by reduced sensitivity to the transverse field $h_x$. Conversely, operation in the non-PD regime does not guarantee such selective filtering: depending on the parameters, it may exhibit comparable or enhanced sensitivity to $h_x$. This selective control over metrological response, dictated by the underlying dynamical phase, constitutes the key insight and practical relevance of our work. The significance of this result lies in showing that complex few-qubit dynamics can be harnessed for quantum sensing. This metrological advantage is experimentally accessible through the CFI of observables such as total magnetization and two-qubit correlations, offering a feasible route to near-term realization.

However, we must also acknowledge the limitations of the present study. Our work serves as a proof-of-concept, and the stability of the PD signatures against environmental decoherence or drive imperfections has not been systematically investigated. Moreover, the selective filtering effect appears to be prominent within specific parameter regions, and a comprehensive mapping of its robustness across the entire phase space remains a task for future work. An important open question concerns how the metrological advantage scales with system size, linking the few-body dynamics studied here to the collective behavior of true many-body time crystals.

Future investigations could therefore proceed in several directions. Including open-system effects will be crucial to assess the protocol's experimental viability, while optimal-control techniques may help identify Floquet protocols that enhance both stability and metrological selectivity. By exploring the functional applications of non-equilibrium dynamics in small, controllable systems, this work contributes a complementary perspective to the broader study of many-body physics, suggesting that even without true long-range order, the rich behaviors of driven systems offer valuable resources for quantum technologies. In particular, ring geometries with periodic boundaries can exhibit a richer structure of dynamical phases beyond the simple PD versus non-PD distinction considered here. Exploring refined phase diagrams in such systems, where distinct phase islands may host different metrological advantages, represents an interesting direction for future work~\cite{PhysRevLett.123.150601}.
\section*{Acknowledgment}
\begin{figure}[b!]
    \centering
    \includegraphics[width=1.0\linewidth]{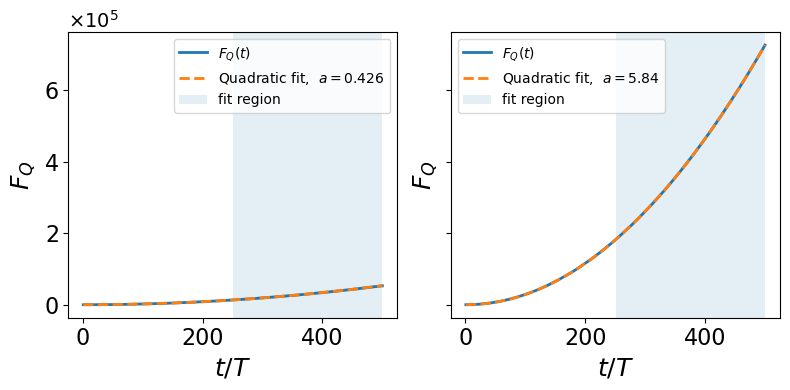}
    \caption{Time evolution of the QFI $F_Q(t)$ for estimating the transverse field $h_x$ at representative points in the PD (left) and non-PD (right) regimes.
The solid lines show the computed $F_Q(t)$, while the dashed lines indicate quadratic fits to the last 50\% of the time series, used to extract the curvature $a$ ($a$ corresponds to the second derivative $d^2 F_Q/d t^2$, i.e., the curvature of the QFI). The shaded regions highlight the portion of the time series used for fitting. The rest of the parameters are the same as in Fig.~\ref{fig:QFI}(a).}
    \label{fig:QFI_fit}
\end{figure}
\begin{figure*}[t!]
    \centering
    \subfloat[]{
    \includegraphics[scale=0.65]{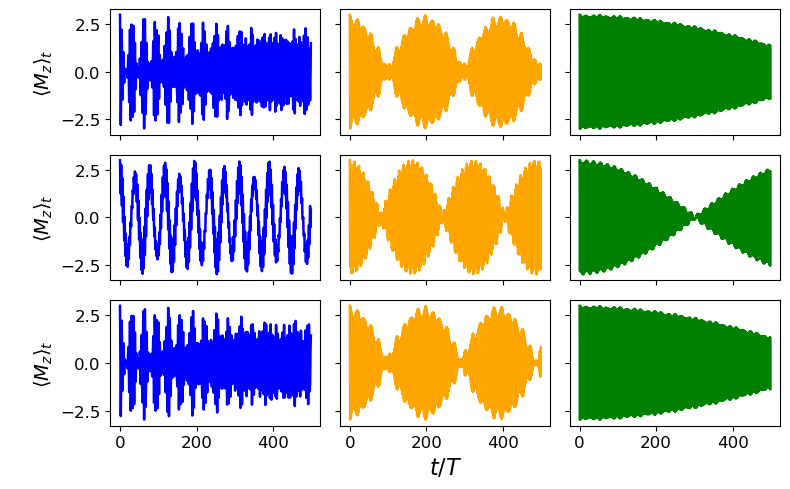}}\\
        \subfloat[]{
    \includegraphics[scale=0.45]{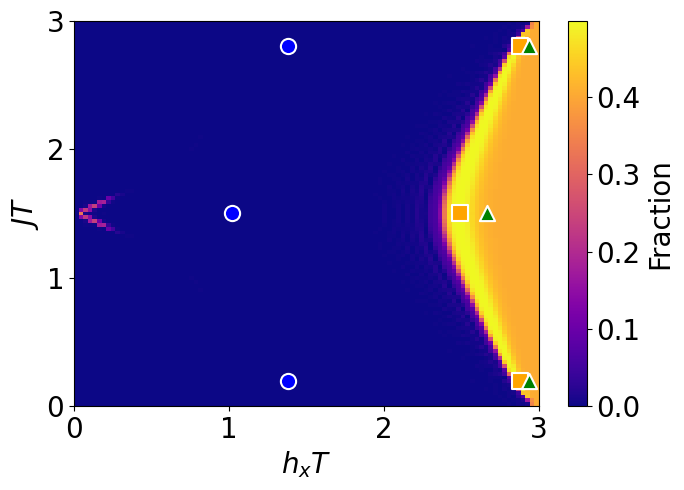}}
    \caption{(a) Time evolution of the magnetization $\langle \hat{M}_z \rangle_t$ for three representative regions of the phase diagram shown in panel (b). Each row corresponds to a fixed value of $J T$ but different values of $h_xT$. The three columns correspond to different points in the phase diagram marked by a circle (left), a square (middle), and a triangle (right). For each $J T$, the three panels represent $h_x T$ values corresponding to the left (blue), middle (orange), and right (green) regions, respectively: 
$J T = 2.94$: $h_x T = 1.445, 3.016, 3.079$; 
$J T = 1.57$: $h_x T = 1.068, 2.608, 2.796$; 
$J T = 0.20$: $h_x T = 1.445, 3.016, 3.079$. 
(b) Phase diagram showing the fraction of PD regions as a function of $h_x T$ and $J T$.  The circular, square, and triangular markers indicate the parameter points corresponding to the time evolutions shown in panel (a). }
    \label{fig:Mag}
\end{figure*}
This work is supported by the Scientific and Technological Research Council (T\"UBİTAK) of T\"urkiye under Project Grant No. 123F150. We also acknowledge funding from the Deutsche Forschungsgemeinschaft (DFG, German Research Foundation) under Project No.~435696605 and through the Research Units FOR~5413/1, Grant No.~465199066. This work was also supported by the QuantERA~II programme (project CoQuaDis, DFG Grant No.~532763411), which has received funding from the EU~H2020 research and innovation programme under GA~No.~101017733. We also acknowledge support from the Leverhulme Trust (Grant No. RPG-2024-112). Further support was received by the ERC grant OPEN-2QS (Grant No.~101164443).
\section*{DATA AVAILABILITY}
The data that support the findings of this article are not publicly available. The data are available from the authors upon reasonable request.

\appendix
\section{QFI Scaling for $N =3$}\label{app:curvature}

In this Appendix, we analyze the QFI curvature to provide a broader view of how metrological enhancement is distributed across the parameter space for both PD and non-PD regimes.
We compute the QFI $F_Q(t)$ for estimating the transverse field $h_x$ at representative points in the PD and non-PD regimes. To extract the asymptotic growth rate, we perform a quadratic fit to the \emph{tail} (last 50\%) of the QFI time series, where transient oscillations have decayed and the long-time behavior dominates.
We fit the data using
\begin{equation}
F_Q(t) = \frac{1}{2}a t^2 + b t + c,
\end{equation}
where $a$, $b$, and $c$ are fitting coefficients.  We take the second derivative of $F_Q$ to find the curvature: $d^2 F_Q/d t^2 = a$, which quantifies the asymptotic growth of the QFI. Figure~\ref{fig:QFI_fit} shows the time evolution of the QFI for estimation of $h_x$ in the PD and non-PD regimes. As expected, the non-PD regime exhibits faster QFI growth, indicating higher sensitivity to $h_x$ at long times. For the PD regime, we find $a = 0.426$, whereas in the non-PD regime $a = 5.84$. This indicates that, in this example, the non-PD regime exhibits faster growth of QFI and therefore higher sensitivity to $h_x$ at long times. The coefficients $a$, $b$, and $c$ are empirical fit parameters that serve to characterize the temporal behavior of QFI rather than representing universal scalings.

\section{Dynamics of different regimes and Floquet-state overlaps}\label{dynamic}

Here, we consider three representative regimes of the phase diagram shown in Fig.~\ref{fig:Mag}(b), corresponding to the left, middle, and right regions, and investigate the corresponding dynamics of $\langle \hat{M}_z \rangle_t$ in Fig.~\ref{fig:Mag}(a). These regimes capture distinct dynamical responses across the PD boundary. From Fig.~\ref{fig:Mag} we can see that in the middle region of Fig.~\ref{fig:Mag}(b), where the subharmonic spectral weight is large and the selected points are shown by orange squares, the magnetization displays pronounced beat patterns: within each short beat, the oscillations occur predominantly at the $2 T$ PD frequency, but the envelope decays relatively quickly. In contrast, the right-hand region (points are shown as green triangles) shows longer beat envelopes, where the oscillation amplitude persists for more cycles, yet the spectral analysis indicates a weaker contribution from the $2 T$ subharmonic. The smaller spectral weight in this region reflects that the dynamics arise from a narrower subset of Floquet quasi-energy pairs and therefore a smaller overlap of the initial state with the $\pi$-paired Floquet subspace. The crescent region, by contrast, involves a larger fraction of $\pi$-paired states and higher initial-state overlap with that subspace, producing a stronger and more collective PD response even though the envelope of the oscillations decays faster. The QFI supports this interpretation: the same region maximizes the QFI for estimating the Ising coupling $J$, consistent with the enhanced sensitivity of pairwise correlations within the $\pi$-paired sector.
\begin{figure}[t]
    \centering
    \includegraphics[scale=0.45]{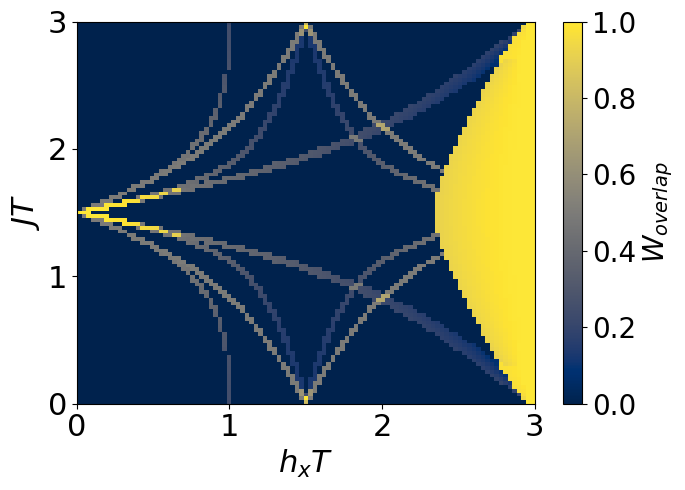}
    \caption{Phase diagram showing the initial state $\ket{\phi_0}$ overlap 
$W_{\mathrm{overlap}}$ [Eq.~\eqref{overlap}] with the $\pi$-paired Floquet states, defined by the condition $|\epsilon_i - \epsilon_j| \approx \pi/T$.  Brighter regions indicate stronger overlap with the $\pi$-paired subspace.}
    \label{fig:overlap}
\end{figure}
The Floquet eigenstates are defined by
\begin{equation}
    U_F \ket{\phi_\alpha} = e^{-i \epsilon_\alpha T} \ket{\phi_\alpha},
\end{equation}
where $\epsilon_\alpha$ are the quasienergies of the system and $T$ is the driving period. 

Two Floquet eigenstates are said to form a \(\pi\)-pair if their quasienergies differ by approximately half of the Floquet zone, such as
\begin{equation}
    |\epsilon_i - \epsilon_j| \approx \frac{\pi}{T}.
\end{equation}
The fraction of such $\pi$-paired states is defined as
\begin{equation}
    f_{\pi} = \frac{N_{\pi}}{N_{\text{tot}}},
\end{equation}
where $N_{\pi}$ is the number of states forming $\pi$-pairs and $N_{\text{tot}}$ is the total number of Floquet eigenstates. To quantify how strongly the initial state $\ket{\psi_0}$ overlaps with the $\pi$-paired subspace, we define the normalized overlap weight, give below
\begin{equation}\label{overlap}
    W_{\text{overlap}} =
    \frac{
        \displaystyle
        \sum_{(\alpha,\beta)\,\pi-paired}
        \left(
            |\braket{\phi_\alpha | \psi_0}|^2 +
            |\braket{\phi_\beta | \psi_0}|^2
        \right)
    }{
        \displaystyle
        \sum_{\gamma} |\braket{\phi_\gamma | \psi_0}|^2
    }.
\end{equation}
The overlap between initial state and Floquet eigenstates is shown in Fig.~\ref{fig:overlap}. A large value of $W_{\text{overlap}}$ indicates that the initial state has significant support on $\pi$-paired Floquet eigenstates, a hallmark of PD dynamics in our system.


\section{QFI curvature for the estimation of $h_x$ and $J$ in larger spin systems}
\label{largerSpins}

\begin{figure*}[t!]
    \centering
    \includegraphics[width=0.9\linewidth]{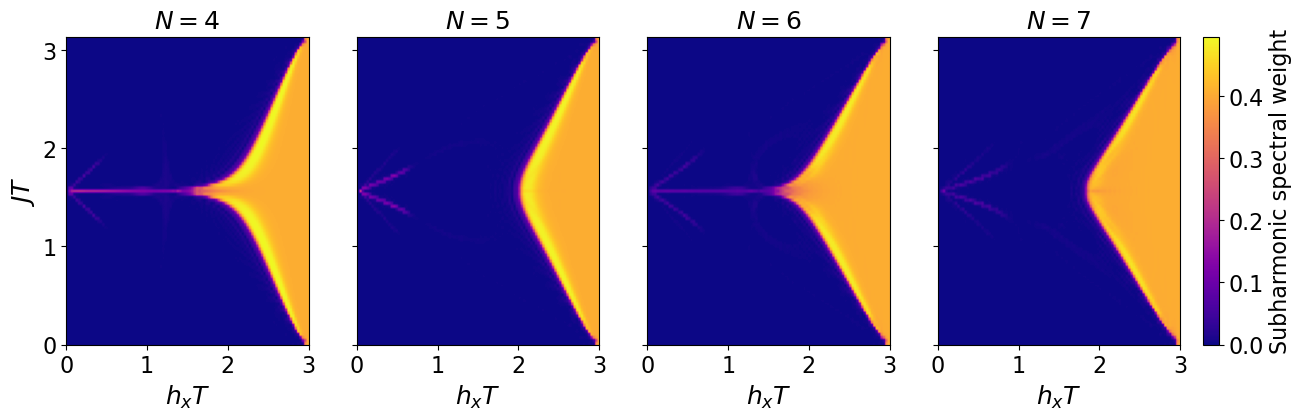}
    \caption{Phase diagrams for different numbers of qubits $N=4,5,6,7$ for PD and non-PD signatures. The blue marks show the PD, and the red cross marks show the non-PD regimes. The rest of the parameters are the same as in Fig.~\ref{fig:Mz+Power+QE}.}
    \label{fig:Phases}
\end{figure*}

To analyze the scaling behavior of metrological performance in larger systems, we extend our study of the QFI curvature, $d^2F_Q/dt^2$, to $N$-qubit Floquet chains. The stroboscopic dynamics is governed by the periodically driven Hamiltonian
\begin{equation}
\hat{H}(t) =
\begin{cases}
\displaystyle \hat{H}_z = J \sum_{i=1}^{N-1} \hat{\sigma}^i_z \hat{\sigma}^{i+1}_z, & \text{for } 0 < t < T_1, \\[8pt]
\displaystyle \hat{H}_x = h_x \sum_{i=1}^{N} \hat{\sigma}^i_x, & \text{for } T_1 < t < T_1 + T_2 = T,
\end{cases}
\label{eq:Hamiltonian_N}
\end{equation}
where $J$ is the nearest-neighbor interaction strength and $h_x$ is the transverse magnetic field. We first map out the parameter regions where these PD regimes occur for larger system sizes. To this end, we compute the relative subharmonic spectral weight for different system sizes $N$ in Fig~\ref{fig:Phases}, and identify the regions of robust PD in parameter space ($h_xT$, $JT$).  The resulting phase diagrams for $N=4,5,6,7$ qubits are shown in 
Fig.~\ref{fig:Phases}.

\begin{figure*}[t!]
    \centering
    \includegraphics[width=0.9\linewidth]{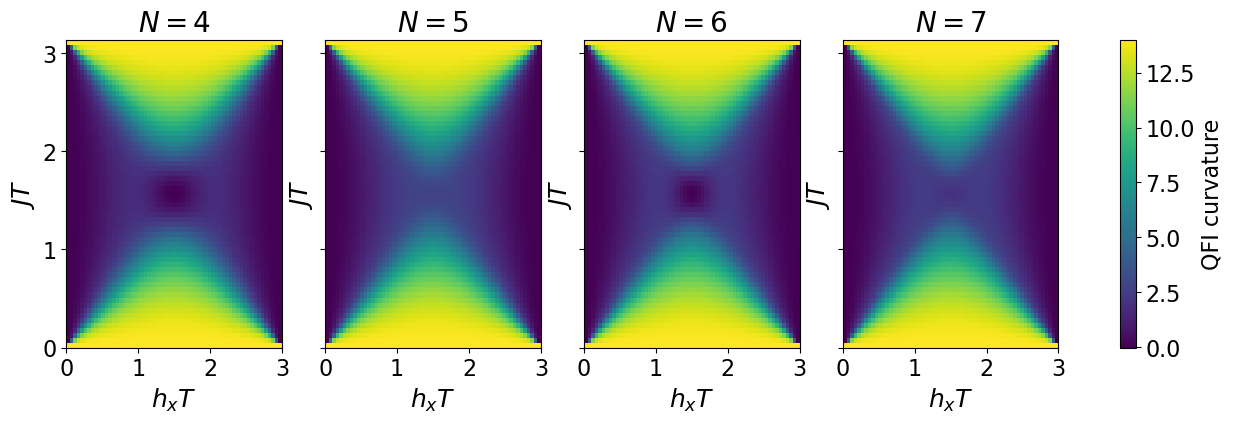}
        \includegraphics[width=0.9\linewidth]{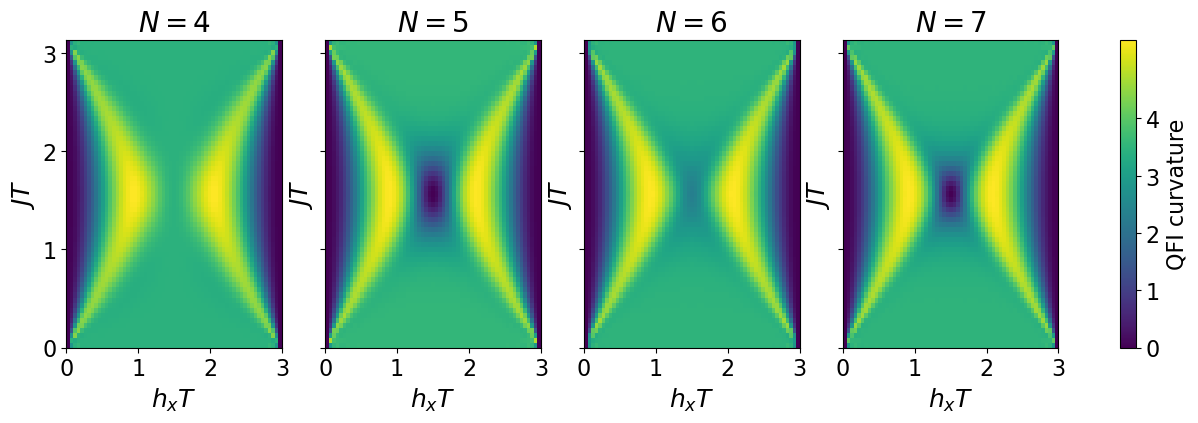}
    \caption{Phase diagrams of QFI curvatures $d^2F_Q/dt^2$ for estimation of magnetic field $h_x$ (top panel) and interaction strength $J$ (bottom panel) as a function of interaction strength $J$ and magnetic field $h_x$ for different system sizes $N$. The rest of the parameters are the same as in Fig.~\ref{fig:Mz+Power+QE}.}
    \label{fig:CurvN4}
\end{figure*}

For completeness, in Fig.~\ref{fig:CurvN4} we extend the QFI curvature analysis [cf.~Eq.~\eqref{QFI_curve}] to larger system sizes $N=4,5,6,7$. The top panels show the curvature of QFI for estimation of $h_x$, while the bottom panels correspond to estimation of $J$. These results demonstrate that the trends observed for $N=3$ persist in larger systems. In particular, the estimation of $h_x$ is consistently enhanced in the non-PD regime, while the PD regime boosts sensitivity to $J$ in larger system sizes. The PD phase is relatively robust with respect to system size $N$ in the $h_xT-JT$ diagram. However, it performs poorly for estimating $h_x$, while offering moderate performance for the estimation of $J$. By contrast, the non-PD phase can reach much higher precision for sensing, though its performance is less predictable: it can yield either very good or very poor estimates depending on the parameter selection regime. We note that the phase diagrams in Fig.~\ref{fig:Phases} exhibit qualitative differences between odd and even system sizes. This even-odd effect is a genuine finite-size phenomenon and persists irrespective of whether Kac scaling is employed, as discussed in Ref.~\cite{Ullah_2026}. For finite $N$, the parity of the system leads to distinct interference patterns in the Floquet dynamics, which manifest as qualitatively different phase diagrams and curvatures of the QFI, as further illustrated in Fig.~\ref{fig:CurvN4}. Physically, these even-odd differences originate from quantum interference~\cite{Schleich1987,PhysRevA.38.1177,PhysRevA.48.1854} and are a well-known feature of finite quantum many-body systems. Similar results have been reported previously in the contexts of quantum metrology~\cite{Alushi2025,Ullah_2026} and quantum heat engines~\cite{PhysRevE.97.042127}.

A notable advantage of the PD phase is that it remains consistently in a high-precision regime, and it never collapses to very low performance. This reliability contrasts with the non-PD phase, whose performance fluctuates. For $h_x$ estimation in particular, the PD phase is consistently unfavorable and thus should be avoided when selecting the operating parameter ($J$) of a magnetometer, especially for sensing magnetic fields over a broad range. Similarly, for the estimation of interaction strength $J$, the magnetic field should be chosen where the PD points are located, such as $h_xT\ge2$ to get high precision. In summary, understanding the characteristics of the PD phase enables more informed choices of sensor operating parameters.
\bibliography{TM}

\end{document}